# Mapping fluorescence enhancement of plasmonic nanorod coupled dye molecules

Emese Tóth[1], Ditta Ungor[2], Tibor Novák[1], Györgyi Ferenc[3], Balázs Bánhelyi[4], Edit Csapó[2,5], Miklós Erdélyi[1], and Mária Csete[1*]

1. Department of Optics and Quantum Electronics, University of Szeged, H-6720, Dóm square 9, Szeged, Hungary
2. Interdisciplinary Excellence Centre, Department of Physical Chemistry and Materials Science, University of Szeged, H-6720 Rerrich B. square 1, Szeged, Hungary
3. Institute of Plant Biology, Biological Research Centre, Hungarian Academy of Sciences, H-6726 Temesvári krt. 62, Szeged, Hungary
4. Department of Computational Optimization, University of Szeged, H-6720 Árpád square 2, Szeged, Hungary
5. MTA-SZTE Biomimetic Systems Research Group, Department of Medical Chemistry, Faculty of Medicine, University of Szeged, H-6720 Dóm square 8, Szeged, Hungary
* Correspondence: mcsete@physx.u-szeged.hu; Tel.: +36-62-544654

**Abstract:** Plasmonically enhanced fluorescence is a widely studied and applied phenomenon, however only a comparative theoretical and experimental analyses of coupled fluorophores and plasmonic nanoresonators makes it possible to uncover, how this phenomenon can be controlled. A numerical optimization method was applied to design configurations that are capable of resulting in an enhancement of excitation and emission, moreover of both phenomena simultaneously in coupled Cy5 dye molecule and gold nanorod systems. Parametric sensitivity studies revealed, how the fluorescence enhancement depends on the molecule's location, distance and orientation. Coupled systems designed for simultaneous improvement exhibited the highest (intermediate directional) total fluorescence enhancement, which is accompanied by intermediate sensitivity to the molecule's parameters, except the location and orientation sensitivity at the excitation wavelength. Gold nanorods with a geometry corresponding to the predicted optimal configurations were synthesized, and DNA strands were used to control the Cy5 dye molecule distance from the nanorod surface via hybridization of the Cy5-labelled oligonucleotide. State-of-the-art dSTORM microscopy was used to accomplish a proof-of-concept experimental demonstration of the theoretically predicted (directional) total fluorescence enhancement. The measured fluorescence enhancement was in good agreement with theoretical predictions, thus providing a complete kit to design and prepare coupled nanosystems exhibiting plasmonically enhanced fluorescence.

**Keywords:** gold nanorod; plasmon resonance; Cy5 dye molecule; enhanced fluorescence; optimization; DNA; STORM

## 1. Introduction

Plasmon enhanced fluorescence phenomenon is a synergetic action of the excitation rate enhancement and the modification of the overall quantum yield, and is influenced by the achievable out-coupling efficiency as well.

In the earliest literature the excitation enhancement was computed based on the **E**-field enhancement around a dipolar emitter, whereas the emission enhancement was treated by a surface induced radiative rate correction [1, 2].

Later the excitation enhancement was simply rescaled by the quantum efficiency modification at the emission, and the antenna collection efficiency was also taken into account. Accordingly, the intensity of a plasmonic antenna coupled emitter was usually computed as I=$\gamma_{exc}$*$\eta_{em}$*$\varepsilon_{coll}$ [3, 4].

It was proven that the involved antenna modes determine the relative strength of radiative and non-radiative rates, and quenching / fluorescence enhancement occurs typically on the blue / red side of the spectral peaks corresponding to resonances on plasmonic nano-objects [5].

Primary studies predicted that the excitation and emission improvement can be compromised, when the single plasmon resonance peak is in between the corresponding wavelengths [6, 7]. On nanorods the co-existent transversal and longitudinal resonances are beneficial, however as a cost of strong polarization dependency [8]. The difference between the *Purcell factor* (total decay rate enhancement) and radiative rate enhancement was attributed to losses in the coupled fluorophore-nanophotonic systems [9]. In our previous studies a novel theoretical approach was developed to extract the total and radiative decay rate enhancements both at the excitation and emission wavelengths [10, 11, 12].

In experimental studies single molecule and single plasmonic nano-object interaction was preferred, since in case of multitudes the plasmonic antennas density has to be larger than a threshold, but their size distribution makes the plasmonic responses spectrally broad and the resulted peaks are inherently blurred caused by the random orientation of fluorophores [13]. A DNA origami was used to position single Cy5 dye molecule and to map the local density of states (LDOS) in plasmonic nanoparticle on mirror type (NPonM) nanoresonators with 1.5 nm resolution. It was shown that coherent coupling of an emitter and a plasmonic nanoresonator results in a modulation of the scattering spectrum [14].

Extended studies on coupled emitter and plasmonic nano-object systems created a great demand for high spatial resolution microscopy techniques. Diffraction–limited resolution can be surpassed for example by stimulated emission depletion (STED) [15], stochastic optical reconstruction microscopy (STORM) [16] and photo-activation localization microscopy (PALM) [17] that are capable of resulting in single-molecule localization (SML) as well.

Three phenomena can improve SML precision in close proximity of metal surfaces: (i) the metal quenches the molecules adsorbed on it that provides a contrast to dielectric objects, (ii) excitation enhancement is also achieved at the (localized) surface plasmon resonances ((L)SPR), (iii) the stability of the molecules is improved, since due to the increased decay the lifetime in excited states is decreased, and the probability that the emitter undergoes photo-induced chemical processes is decreased as well [9, 18].

Accordingly, resolution enhancement was achieved via surface plasmon resonance illumination [19]. Novel microscopy methods appeared as well, e.g. Brownian motion single molecule super-resolution imaging [20], moreover several former microscopy techniques were developed by using plasmonic nanoparticles in e.g. in STED [21].

It was shown that the apparent position of a dipolar emitter in close proximity of a plasmonic antenna is governed by a distance dependent partial coupling due to the decay of the molecule through two competing radiative pathways: directly and mediated by the antenna [22]. Moreover, it was described that the apparent position of emitters coupled to metal nanorods can be affected by the objective induced point-spread function aberrations [23]. After excluding the effect of dye concentration, quality, molecule distance and image dipole formation the apparent smaller nanorod size was attributed to heterogeneous binding of double-strand DNA used as a linker [24]. This indicates that there exists a great demand for predesigned DNA linkers' development.

In case of STORM the advantage of controlled emission detuning from SPR via a large Stokes shift was proven, since it promoted large localization precision and high-resolution electromagnetic-field mapping, whereas fluorophores coupled at their emission made it possible to explore the LDOS distribution with high-resolution [25].

The synergy between super-resolution imaging and plasmonics can result in significant resolution improvement, accordingly plasmonic NPs can be used as image contrast agents, whereas plasmonically tailored excitation fields can promote sub-diffraction-limited spatial resolution, e.g. in STED [26]. The optimal spectral and spatial conditions to improve super-resolution imaging of plasmonic particles correspond to a partial decoupling, which makes it possible to map the realistic nano-object shape [27, 28].

This has been demonstrated via motion-blur point accumulation for imaging in nanoscale tomography (PAINT) and via DNA-PAINT as well. However, systematic microscopy on a single NP exhibiting resonance at a spectral location exactly in between the excitation and emission wavelength has not been reported previously.

In our present study coupled Cy5 dye molecule and gold nanorod (Au NR) systems have been designed via numerical computations to improve excitation and emission simultaneously. Analogue Cy5 - Au NR systems were prepared by controlling their distance via double-strand DNA and then inspected via dSTORM. The fluorescence responses detected by dSTORM have been compared to the (directional) fluorescence enhancement that is achievable via coupled Cy5 dye molecule and gold nanorod configurations optimized by special numerical computation methods.

## 2. Methods

*2.1. FEM optimization and analyses of optimal configurations*

In our previous studies we have shown that the radiative rate enhancement ($\delta R$) of a dipolar emitter can be determined by multiplying the *Purcell factor* (which is the total decay rate enhancement in a plasmonic nanoresonator with respect to a homogeneous environment) and the quantum efficiency (*QE*, which is the ratio of the power radiated into the far-field to the total emitted power) at the wavelength either of excitation or emission:

$$\textit{Purcell factor}= P_{total}/P_{total\_0}= (P_{radiative}+P_{non-radiative}) /(P_{radiative\_0}+P_{non-radiative\_0}) \quad (1)$$

$$QE= P_{radiative}/(P_{radiative}+P_{non-radiative}) \quad (2)$$

$$\delta R= \textit{Purcell factor}*QE \quad (3)$$

In coupled systems consisting of a fluorophore and a plasmonic nanoparticle a significant part of the emitted power is transferred to the NP, however only a fraction of it escapes to the far-field, whereas the remaining part is lost in form of a resistive heating. This explains, why optimization is necessary to determine configurations that are the most appropriate for fluorescence enhancement.

2.1.1. Optimization of Cy5 dye molecule and gold nanorod coupled systems

The excitation (635 nm) and emission (665 nm) wavelengths of the inspected Cy5 dye molecule are close to each other, so it is expected that both phenomena can be enhanced by a single broad dipolar plasmonic resonance of a predesigned gold nanorod. Accordingly, the *Purcell factor*, quantum efficiency and radiative rate enhancement of the coupled systems have been monitored at both wavelengths. The intrinsic quantum efficiency (~25%) of Cy5 dye molecule is low, therefore *QE* improvement may be expected as well. Three sets of optimizations have been performed by applying the following objective functions and the $QE_{em\_crit}=25\pm0.25\%$ criterion regarding the quantum efficiency that have to be met at the emission: (i) radiative rate enhancement at the excitation ($\delta R_{exc}$), (ii) radiative rate enhancement at the excitation and emission simultaneously ($\delta R_{[exc,em]}$), (iii) radiative rate enhancement at the emission ($\delta R_{em}$).

The modification of the directivity ($\delta D=D/D_0$, where $D$ and $D_0$ is the directivity of the coupled Cy5 – Au NR system and uncoupled Cy5 dye molecule, respectively) has been evaluated to determine the fraction of enhanced emission ($\delta D*\delta R$) that could be detected from the substrate side at either wavelengths. At the excitation the directivity was computed by supposing an exciting beam that is perpendicularly incident onto the substrate and is polarized along the x axis, i.e. along the long axis of the gold nanorod.

At the emission outgoing beams into all directions were taken into account, independently of polarization. Finally, the total fluorescence enhancement, which is qualified by the $P_x$ *factor* = $\delta R_{exc}*\delta R_{em}$ as well as the fraction of the total fluorescence enhancement that can be captured via dSTORM microscopy ($D_x$ *factor*= $P_x$ *factor* $\delta D_{exc}$ $\delta D_{em}$) has been evaluated for all of the optimized systems (Table 2).

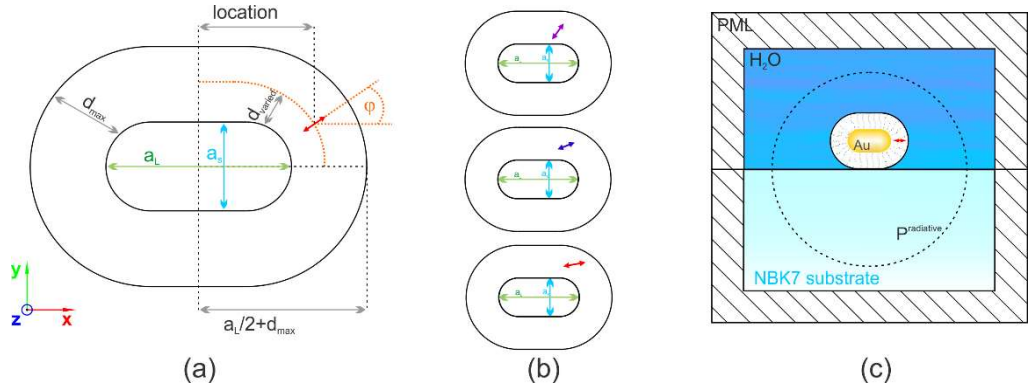

**Figure 1.** (a) Schematic drawing of the coupled system parameter's variation: *dipole location* (x) along the long axis of and *distance* (d) from the nanorod, as well as *orientation* (φ) with respect to the long axis, (b) schematics of the configurations received via i-ii-iii type optimizations, (c) extraction methodology of quantities used to evaluate the coupled Cy5 dye molecule – gold nanorod systems

The geometry of the optimized gold nanorods has been qualified by their short ($a_s$) and long ($a_l$) axis, based on these parameters the aspect ratio (*AR*) has been also computed. These parameters were varied in the interval of short and long axis of nanorods that were determined in preliminary TEM measurements (Fig. 1, 5, Table 2). Additionally, the following parameters of dipolar emitters corresponding to Cy5 dye molecules have been varied: dipole location *(x)*, qualifying the distance from the center along the long axis of the nanorod; dipole distance *(d)*, measured perpendicularly from the nanorod surface inside a DNA strand having a fixed thickness of 18 nm; dipole orientation *(φ)*, measured from the long axis of the nanorod (Fig. 1a).

2.1.2. Analyses of optimal Cy5 dye molecule and gold nanorod coupled systems

The wavelength dependence of the *Purcell factor*, quantum efficiency and radiative rate enhancement has been determined for the optimal configurations received in i-ii-iii optimization cases (Fig. 1b, 2a-c). In these computations the wavelength of the dipolar emitter modeling the Cy5 dye molecule has been tuned, and the model consisting of a single Cy5 dye molecule and a single gold nanorod coupled system on a substrate has been terminated by a PML layer (Fig. 1c). The charge and near-field distribution (Fig. 3a-c) as well as the far-field radiation pattern (Fig. 3d-f and g-i) have been analyzed in the optimal configurations of i-ii-iii type coupled systems as well.

2.1.3. Sensitivity study on Cy5 dye molecule and gold nanorod coupled systems

It was studied, how the achievable *Purcell factor*, *QE* and radiative rate enhancement varies, when the location of the Cy5 dye molecule is tuned from the center through the apex of the gold nanorod at the optimal distance and orientation (Fig. 4a-c), and when the distance of the Cy5 dye molecule from the gold nanorod surface is increased at the optimal location and orientation (Fig. 4d-f). The orientation dependence of the *Purcell factor*, *QE* and radiative rate enhancement has been inspected as well at the optimal location and distance of the Cy5 dye molecule inside the i-ii-iii type coupled systems (Fig. 4g-i).

In our present study dipolar emitters aligned in the xy plane have been analyzed, however similar behavior is expected in xz and yz plane, and an additional comparative study on them will be presented in an upcoming theoretical work.

*2.2. Preparation protocol and characterization of Au NRs*

2.2.1. Materials used to prepare Au NRs

All chemicals were of analytical grade and were used without further purification. The gold (III) chloride acid trihydrate (HAuCl$_4$×3H$_2$O; ≥99.9 %) and the hexadecyltrimethylammonium bromide (C$_{16}$H$_{33}$N(CH$_3$)$_3$Br, ≥98 %) were from Sigma. The hydrogen chloride solution (HCl, 37%), the sodium borohydride (NaBH$_4$; 99 %) and the silver nitrate (AgNO$_3$; ≥98 %) were from Molar.

2.2.2. Preparation protocol of Au NRs

The Au NRs were synthesized by a modified soft template method [29]. Firstly, the seed solution was prepared by adding 250 μL 0.01 M HAuCl$_4$ to 10.0 mL 0.1 M CTAB solution. Upon the appearance of the dark orange color, freshly prepared ice-cold NaBH$_4$ (0.6 mL 0.01 M) solution was quickly added into the mixture and stirred for 5 min. The formation of the brown color refers to the formation of gold seeds in the solution. The gold colloids were kept at room temperature for at least 2 hours before using them. For the seed-mediated growth 2.0 mL 0.01 M HAuCl$_4$ and the 150 μL of 0.01 M AgNO$_3$ were compounded with 40 mL 0.1 M CTAB solution. After the mixing 0.32 mL 0.1 M freshly prepared ascorbic acid, 0.8 mL 1M HCl and 96 μL seed solution were added into the solution. The reaction mixture was blended by gentle shaking. Finally, the growing solution was left undisturbed for 24 h at 40 °C.

2.2.3. Characterization of the synthesized Au NRs

For the optical characterization, the UV-Vis-NIR spectra were recorded on JASCO, V-770 Spectrophotometer using 1 cm quartz cuvette from 190 nm to 1000 nm. High resolution transmission electron microscopy (HRTEM) images were captured on a FEI Technai G2 instrument at 200 kV accelerating voltage. The samples were dried on carbon film Cu grids (Electron Microscopy Sciences, 200 mesh). The images were analyzed by the ImageJ software. The aspect ratio of the synthesized Au NRs was determined based on TEM images and on UV-Vis spectra [30]. (Fig. 1, 5).

2.3. Synthesis and purification of thiol- and Cy5-modified DNA oligonucleotides, functionalization of Au NRs with thiol-modified DNA oligonucleotide and hybridization of the Cy5-labelled oligonucleotide

2.3.1. Materials used to create DNA strands

An in-house synthesized 4,4′,4″-trimethoxytrityl (TMTr)-protected 5′-Thiol-Modifier C6 phosphoramidite was used [31]. Phosphoramidites, synthesis reagents and solvents used for DNA oligonucleotide synthesis were purchased from Sigma, Molar and Linktech.

2.3.2. Synthesis and purification of thiol and Cy5-modified DNA oligonucleotides

DNA oligonucleotides were synthesized using a DNA/RNA/LNA H-16 synthesizer (K&A Laborgeraete) by standard β-cyanoethyl phosphoramidite chemistry at a nominal scale of 0.2 μmol. Sequences of the complement oligonucleotides are included into Table 1.

Oligonucleotides were purified by HPLC on an RP-C18 column under ion-pairing conditions (using mobile phases containing 0.1 mM triethylammonium acetate (TEAA), pH 6.5 buffer). The 4,4′-dimethoxytrityl and 4,4′,4″-trimethoxytrityl protective groups from the 5′-end of the oligonucleotide were removed using 2% TFA on a Poly-Pak column. After lyophilisation, 100 μM stock solution was prepared from the oligonucleotides.

2.3.3. Functionalization of Au NRs with thiol-modified DNA oligonucleotide

Functionalization of Au NRs with thiol-modified-DNA was done based on mPEG-SH/Tween 20-assisted method in order to avoid aggregation of the positively charged Au NRs and negatively charged DNA [32]. Using this fast method, in 1 h Tween 20 and mPEG-SH displaced CTAB on the surface of Au NRs and 5′-thiol-modified-DNA can be attached to it in 100 mM solution of sodium-citrate.

Accordingly, Au NRs were functionalized based on Jiuxing Li et al. method with the following modifications: (1) mPEG-SH was added two times [31]; (2) after 1 h aging DNA-Au NRs were washed two times with 20 mM HEPES buffer (pH 7.1) and after final centrifugation, resuspended in 20 mM HEPES (pH 7.1) with 50 mM NaCl (100 μL).

**Table 1.** TMTr-Thiol-ModifierC6 (SH-C6) and Cy5-fluorescent label were coupled by phosphoramidite chemistry to the 5′-end of oligonucleotides on solid-phase.

| | |
|---|---|
| thiol-DNA | SH-C6-AATCTGTATCTATATTCATCATAGGAAACACCAAAGATGATATTTTCTTTAAT |
| Cy5-DNA 17.5 nm | **Cy5**-ATT AAAGAAAATATCATCTTTGGTGTTTCCTATGATGAATATAGATACAGATT |
| Cy5-DNA 8.9 nm | ATTAAAGAAAATATCATCTTTGGTGT+**Cy5**-TTCCTATGATGAATATAGATACAGATT |

2.3.4. Hybridization of the Cy5-labelled oligonucleotide to the DNA-Au NR

In the following step a 10 μL solution of Cy5-labelled complement DNA (in 100 μM concentration) was added to the 100 μL DNA-Au NR solution in 20 mM HEPES with 50 mM NaCl, and incubated at 65 °C for 5 min, then gradually brought to room temperature (−2 °C/min). Unbound oligonucleotide was removed by centrifugation and Au NR functionalized with Cy5-labelled DNA duplexes was resuspended in 20 mM HEPES (pH 7.1) and it was stored at 4 °C until further use [33]. Several Cy5 dye molecule distances in the interval of [0 nm, 18 nm] were defined inside DNA strands having a thickness of 18 nm. In microscopy measurements detailed in the present paper Cy5 dye molecule distance of 8.9 nm has been applied.

*2.4. dSTORM optical system, sample preparation, measurement and analysis*

2.4.1. dSTORM optical system

Super-resolution dSTORM measurements were performed on a custom-made inverted microscope based on a Nikon Eclipse Ti-E frame. After being conditioned (through spatial filtering via fiber coupling and beam expansion) the applied laser beams were focused onto the back focal plane of the microscope objective (Nikon CFI Apo 100x, NA=1.49), which produced a collimated beam on the sample. All the dSTORM images were captured with linearly polarized beam and EPI illumination at an excitation wavelength of 647 nm (MPB Communications Inc.: 647 nm, $P_{max}$=300 mW). The laser intensity was controlled via an acousto-optic tunable filter (AOTF). Images were captured by an Andor iXon3 897 EMCCD camera (512x512 pixels with 16 μm pixel size). Frame stacks for dSTORM super-resolution imaging were typically captured at a reduced image size (crop mode). Excitation and emission wavelengths were spectrally separated with a fluorescence filter set (Semrock, LF405/488/561/635-A-000) and using an additional emission filter (Semrock, BLP01-647R-25) in the detector arm. During the measurements, the perfect focus system of the microscope was used to keep the sample in focus with a precision of <30 nm.

2.4.2. dSTORM sample preparation

Cover slips were plasma treated and coated with PAH (poly(allylamine hydrochloride)) cationic polyelectrolyte monolayer. The solution of the functionalized gold nanorods was dropped on the cleaned glass cover slips.

After five minutes of sedimentation time, the cover slip was rinsed with pure water (HiPerSolv CHROMANORM® for HPLC, VWR Chemicals BDH®) to reduce the number of unbound oligomers. Right before the measurement, the water was replaced with GLOX switching buffer [34] and the sample was mounted onto a microscope slide (Fig. 6a).

2.4.3. dSTORM measurement

Before the dSTORM image acquisition, the positions of Au nanorods were identified with cross-polarized white light illumination. Then the selected areas were checked via fluorescence mode using low excitation power. This procedure ensured that the selected areas contained labeled nanorods as both the scattered light signal of the plasmonic nanoparticles and the fluorescence signal of the dye molecules were detected. For the dSTORM measurement, the excitation intensity was raised to 3 kW/cm$^2$, and 2X10,000 frames were captured with a cropped area size and an exposure time of 105*105 pixels and 30 ms, respectively.

2.4.4. dSTORM analysis

The captured and stored image stacks were evaluated and analyzed with the rainSTORM localization software [35] that fitted a Gaussian point spread function to the images of individual fluorescent Cy5 dye molecules. Localizations belonging to the same blinking event were concatenated with the trajectory fitting algorithm of rainSTORM. The acquired trajectories were used to determine the peak intensities of the PSFs of the blinking events by calculating the average of the intermediate localizations of the given trajectory. Therefore, all trajectories shorter than three frames were automatically discarded. The background of this strong restriction is based on the assumption that the ON-state lifetime of the enhanced fluorophores is increased, since the S1->S0 fluorescence path is getting more preferable than the S1->T1 relaxation [36]. Since in the dSTORM method the molecules can be turned off via their T1 state [34], their ON-state lifetime must be increased (Fig. 7).

**3. Results**

*3.1. Optimized coupled Cy5-Au NR configurations*

3.1.1. Optical response of the optimized configurations

The geometrical parameters are similar in the i-ii-iii type optimized coupled systems (Table 2). The optimal $a_l$ long axis gradually increases by increasing the wavelength, where the radiative rate enhancement was maximized. Although, the optimal $a_s$ short axis increases slightly as well, the optimal *AR* aspect ratio increases gradually, these tendencies altogether ensure the forward shifting of the resonance wavelength to the desired wavelength.

The optimal location of plasmonically enhanced Cy5 dye molecule corresponds to a gradually increasing $x$ coordinate, the distance from the nanorod increases as well, at the same time the optimal angle with respect to the long axis gradually decreases. These tendencies are in accordance with intuitive expectations based on previously described plasmon enhanced emission phenomena in similar coupled systems. All geometrical parameters take on an intermediate value in the system optimized to enhance the excitation and emission phenomena simultaneously.

The optimization performed to enhance excitation / excitation and emission simultaneously / emission resulted in a coupled Cy5 dye molecule and gold nanorod system, which exhibits a local maximum on the *Purcell factor* slightly below the desired wavelength, while the *QE* quantum efficiency shows an increasing tendency, as a result on the $\delta R$ radiative rate enhancement spectrum a global maximum appears at 635 nm / 650 nm / 660 nm (Fig. 2a / b / c).

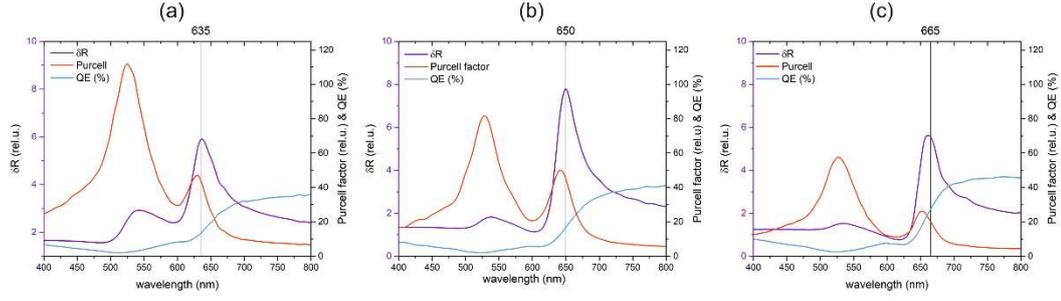

**Figure 2.** *Purcell factor*, quantum efficiency (*QE*) and radiative rate enhancement (δ*R*) spectra of the coupled systems optimized to maximize (a) excitation, (b) excitation and emission simultaneously and (c) emission of Cy5 dye molecule.

**Table 2.** Optical responses at excitation and emission wavelength and geometrical parameters of the optimized systems.

|  | excitation (635 nm) | | medial (650 nm) | | emission (665 nm) | |
|---|---|---|---|---|---|---|
|  | 635 nm | 665 nm | 635 nm | 665 nm | 635 nm | 665 nm |
| *QE* (%) | 13.4 | 25.8 | 9.7 | 24.75 | 8.4 | 27.5 |
| *Purcell factor* | 44.4 | 15.4 | 47.8 | 24.2 | 14.7 | 19.8 |
| δ$R$ | 6.0 | 4.0 | 4.6 | 6.0 | 1.2 | 5.4 |
| δ$D=D/D_0$ | 1.4 | 0.7 | 0.7 | 0.7 | 0.5 | 0.7 |
| δ$D$ *δ$R$ | 8.3 | 2.7 | 3.4 | 4.1 | 0.7 | 3.7 |
| $P_x$ | 23.6 | | 27.6 | | 6.7 | |
| $P_x$*δ$D_{exc}$*δ$D_{em}$ | 22.5 | | 13.8 | | 2.4 | |
| $\eta_{outcoupling}$ | 0.96 | | 0.50 | | 0.36 | |
| $a_S$(nm) | 20.1 | | 20.2 | | 21.9 | |
| $a_L$(nm) | 37.7 | | 39.8 | | 44.6 | |
| AR | 1.87 | | 1.97 | | 2.04 | |
| location "x" (nm) | 9.4 | | 16.6 | | 19.3 | |
| distance "d" (nm) | 9.5 | | 9.7 | | 10.2 | |
| orientation $\varphi$ (°) | 53.0 | | 21.7 | | 10.3 | |

The total fluorescence enhancement achievable in i-ii-iii type optimized coupled systems is in the order of 10. The $P_x$ factor exhibits a maximum in the case of ii-type optimized coupled system proving that it is reasonable to consider it as a representative optimal configuration capable of enhancing both phenomena simultaneously. Considering the total fluorescence enhancement, the optimization performed at the medium wavelength is capable of resulting in a coupled system, where the two phenomena are simultaneously maximized rather than only compromised, in contrast to the literature (Fig. 2b, Table 2) [6, 7].

Taking into account the directivity modification in case of Cy5 dye molecules lying in the xy plane, the fraction of the total fluorescence enhancement that could be collected from the substrate side is approximately 96% / 50% / 36% of the total fluorescence enhancement. The gradually decreasing $D_x$ factor quantifying the directional total fluorescence enhancement indicates that the out-coupling efficiency decreases, when the target wavelength increases.

All optimized coupled systems result in a local field enhancement via dipolar modes on the gold nanorods both at the excitation and emission wavelength (Fig. 3a-c).

The coupled system optimized to enhance the excitation / emission phenomenon exhibits stronger local field enhancement and larger far-field lobes at the excitation / emission wavelength, whereas in the system optimized to enhance these phenomena simultaneously commensurate lobes prove simultaneous intermediate enhancement at both wavelengths in accordance with the expectations (Fig 3 a, d, g/ b, e, h / c, f, i).

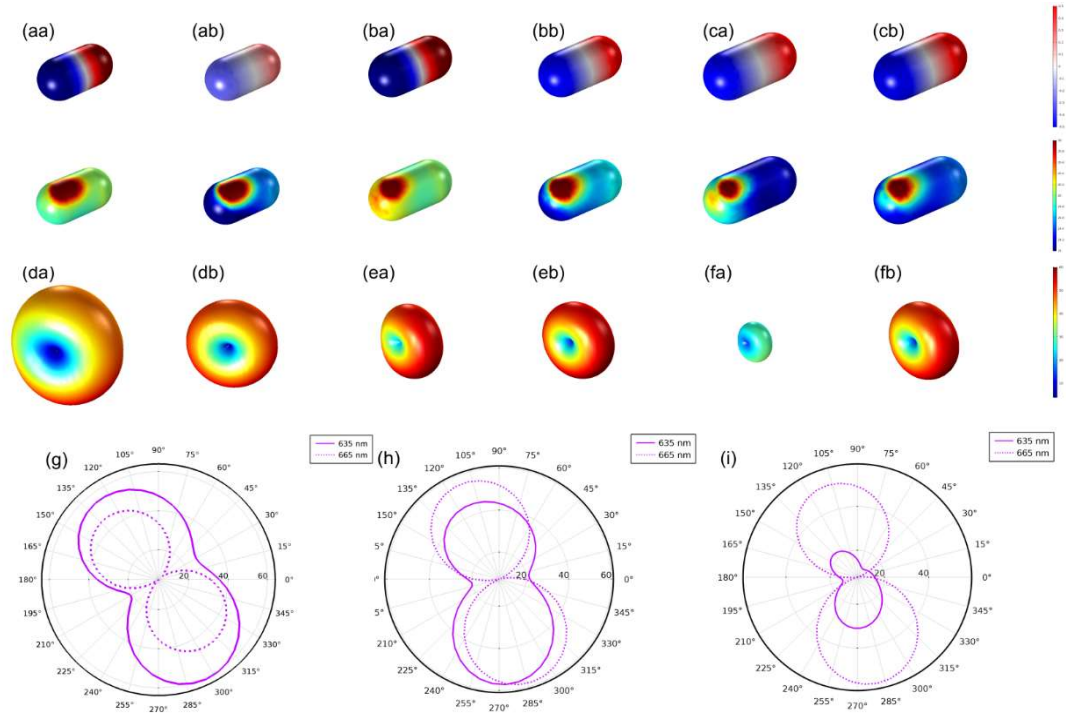

**Figure 3.** Near-field and far-field response of the optimized systems: (a-c) normalized **E**-field and charge distribution, (d-f) far-field radiation pattern at (a-f/a) excitation and (a-f/b) emission wavelength, (g-i) polar diagram at the excitation and emission wavelengths (continuous and dashed line) of the system optimized to enhance (a, d, g) excitation, (b, e, h) excitation and emission simultaneously and (c, f, i) emission of Cy5 dye molecule.

3.1.2. Optical response dependence on dipole location, distance and orientation

A sensitivity study uncovered, how the optical response is influenced in realistic systems, where the distance of the Cy5 dye molecule can be tailored via the presently developed method of Cy5-labelled oligonucleotide hybridization to the DNA-Au NR. One has to note that the dipole location would be hard to control experimentally, and the coupled Cy5 dye molecules orientation is random. However, by comparing computed and experimentally achieved enhancement values it is possible to consider, whether the realistic coupled system corresponds with the predesigned one.

The dipole location, distance and orientation was varied sequentially with respect to the optimal ($x_{opt}$ location, $d_{opt}$ distance, $\varphi_{opt}$ orientation) parameter set. By varying the location of the Cy5 dye molecule, namely by increasing the distance from the center of the gold nanorod all of the *Purcell factor*, *QE* and radiative rate enhancement non-monotonously vary (Fig. 4a-c). At the optimal distance the *QE* meets the criterion set at the emission, and both the *Purcell factor* and the radiative rate enhancement exhibits an increasing tendency. In case of the coupled system optimized to enhance the excitation rate smaller distance from the nanorod center could result in larger *QE*, but the achieved δR would be smaller at both wavelengths (Fig. 4a).

Similarly, in case of the coupled system optimized to enhance excitation and emission simultaneously the *QE* is decreased to the criterion by following a maximum close to the optimal dipole location, whereas the *Purcell factor* and δR exhibits an increasing tendency (Fig. 4b).

In case of the system optimized to enhance the emission the *QE* exhibits a maximum exactly at the optimal location at the emission wavelength, as a consequence either smaller or larger distance from the center would result in smaller *QE* than the criterion.

This indicates that it is not advantageous to tune the location, even though δ*R* would be larger at distances smaller than 15 nm / 7.5 nm at the excitation / emission wavelength (Fig. 4c). Accordingly, the optimal dipole location corresponds to compromised *QE* and δ*R* in all optimized coupled systems. The sensitivity study on dipole location predicts that the achievable radiative rate enhancement varies in (30.40) 12.44 / (31.04) 27.39 / (18.86) 50.00 interval, when the Cy5 dye molecule with optimal distance and orientation is moved from the Au NR center towards the tip. The ii-type coupled system is (the most) intermediately sensitive to the Cy5 dye molecule location at the (excitation) emission wavelength.

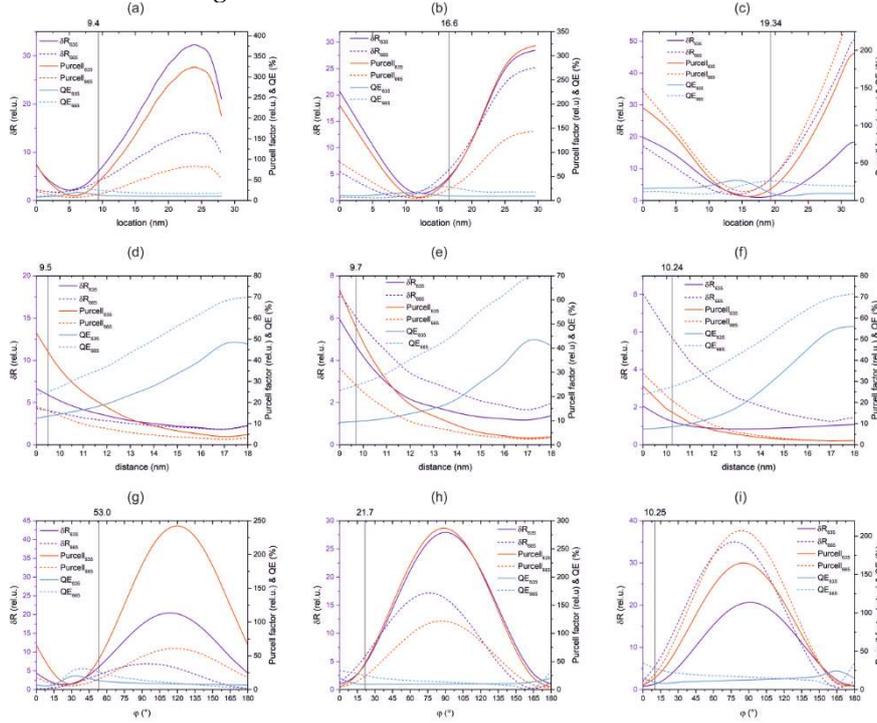

**Figure 4.** Dependence of the achievable *Purcell factor*, quantum efficiency (*QE*) and radiative rate enhancement (δ*R*=*Purcell*\**QE*) on the (a-c) dipole location, (d-f) distance from the nanorod inside 18 nm thick DNS strand, (g-i) orientation of the Cy5 dye molecule, whereas the other two parameters are constant in the ($x_{opt}$ location, $d_{opt}$ distance, $\varphi_{opt}$ orientation) parameter set in case of coupled systems optimized to enhance (a, d, g) excitation, (b, e, h) excitation and emission simultaneously and (c, f, i) emission phenomenon of Cy5 dye molecule.

By varying the distance of the Cy5 dye molecule from the nanorod surface inside the 18 nm thick DNS strand both the *Purcell factor* and the radiative rate enhancement monotonously decreases, whereas the quantum efficiency exhibits a maximum at larger distance in the coupled system optimized to enhance both phenomena / monotonously increases at both wavelengths in the systems optimized to maximize the radiative rate at 635 nm and 665 nm. The optimal distance is the threshold, above which the *QE* meets the criterion set at the emission (Fig. 4d-f).

The sensitivity study on dipole distance predicts that the achievable radiative rate enhancement varies in (4.81) 2.50 / (4.76) 5.5 / (1.22) 6.81 interval at the (excitation) emission wavelength, when the Cy5 dye molecule with optimal location and orientation is moved from the Au-NR surface towards the top of the DNS strand. The ii-type coupled system is intermediately sensitive to the Cy5 dye molecule distance at both wavelengths.

By varying the orientation of the Cy5 dye molecule with respect to the long axis of the nanorod all of the *Purcell factor*, *QE* and radiative rate enhancement non-monotonously vary. At the optimal orientation the *QE* meets the criterion set at the emission wavelength, and both the *Purcell factor* and the radiative rate enhancement exhibits an increasing tendency.

Although, smaller tilting with respect to the long axis could result in larger *QE*, but the achieved δ*R* would be smaller. Accordingly, the optimal orientation corresponds to compromised *QE* and δ*R* (Fig. 4 g-i).

The sensitivity study on dipole orientation predicts that the achievable radiative rate enhancement varies in (18.61) 6.55 / (26.94) 16.94 / (19.91) 34.45 interval, when the Cy5 dye molecule with optimal location and distance is rotated. The ii-type coupled system exhibits (the highest) intermediate sensitivity to the Cy5 dye molecule orientation at the (excitation) emission wavelength.

*3.2. Size distribution and spectral properties of gold nanorods*

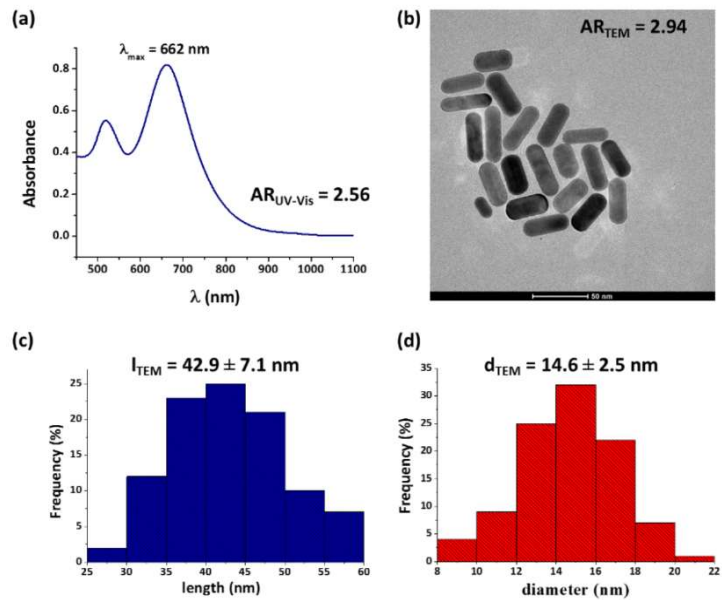

**Figure 5.** The UV-Vis spectrum (a) and the TEM image (b) of the chemically synthesized Au NRs. The long (c) and short (d) axes distributions of the prepared gold nanorods calculated based on TEM images.

Based on TEM images the mean value of the gold nanorods short and long axis is 14.6 nm ± 2.5 nm and 42.9 nm ± 7.1 nm, respectively. This corresponds to the mean aspect ratio of 2.94, which is close to the *AR* predicted by optimizations. Two peaks are observable on the spectra of the nanorods, the local and global maximum at ~520 nm and 662 nm corresponds to the transversal and longitudinal resonance of the nanorod, respectively (Fig. 5).

The transversal resonance is almost coincident with the particle plasmon resonance of gold according to the small short nanorod axes, whereas the longitudinal resonance indicates that the nanorods act as λ/2 mode antennas at 662 nm. The co-existence of resonance nearby the excitation and emission wavelengths of Cy5 dye molecule is capable of resulting in LDOS enhancement and in the improvement of both phenomena.

*3.3. dSTORM imaging of Au nanorods labelled with Cy5 dye molecule*

The surface strands were bonded to the Au nanorods with high specificity via the thiol group. After adding the probe strands to the sample, they predominantly bonded to the surface strands and after their hybridization they formed a stable double strand structure.

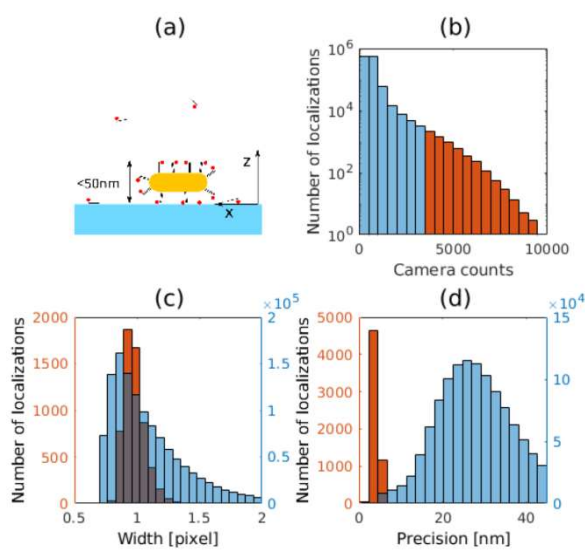

**Figure 6.** (a) Schematic figure of a functionalized Au nanorod on the coverslip. (b) Localizations with a brightness value higher than 3500 camera counts were selected for further evaluations. The intense localizations correspond (c) to the in focus localizations with FWHM≈1pixel and (d) to the most precisely (<6 nm) localized ones. Localizations marked with orange in Figures (b), (c) and (d) belong to the same group. Please note that Cy5 dye molecule labelled DNS can be found in each plane cross-section in reality.

It was assumed that during the STORM measurements the Au nanorods lay parallel to the surface of the cover slip and are stable (Figure 6a). Despite this very high specificity, a small amount of probe strands either attached to the surface of the cover slip or moved freely in the buffer providing positive false localizations, and increased the background of the STORM image. The accepted localizations were arranged based on their fitted Gaussian peaks and only the most intense ones with peak values larger than 3500 counts were selected for further evaluations (Figure 6b). Such a threshold was set as it is higher than the double value of the average peak intensity of individual, not bonded dye molecules. Therefore, we can assume that the filtered bright spots were the images of fluorescence enhanced Cy5 dye molecules bounded to the Au nanorods and the overlapping but unenhanced localizations were discarded. The histogram of the FWHMs of the fitted Gaussian PSF (Figure 6c) and the Thompson localization precisions (Figure 6d) prove that the selected bright spots correspond to the most perfectly focused and most precisely (<6 nm) localized molecules, respectively [37].

Individual gold nanorods were identified by DBSCAN cluster analysis of the accepted localizations [38]. This algorithm requires two input parameters: a minimum number of points that forms a cluster ($N$) and the maximum distance between two adjacent points ($\varepsilon$). Clusters with $\varepsilon$=160 nm and $N$=8 set values were considered as footprints of plasmonically enhanced fluorescent Cy5 dye molecules on gold nanorods. Figures 7a and 7b depict two sample dSTORM images of isolated nanorods (no additional clusters in the neighboring 320×320 nm² area).

Each of these nanorods features two hot spots separated by 30-40 nm, which were identified as the two apexes of the nanorods. Figures 7c and 7d show the histogram of the mean and maximum fluorescence enhancement values for more than 260 selected nanorods. While the maximum value of enhancement was found to be 20 and shows an exponential decay, the mean value was typically below 10. We would like to emphasize that the Cy5 dye molecule fluorescence emission is enhanced by the plasmonic resonance of the gold nanorod and exhibits a well-defined directivity in contrast to the rotary photoluminescence of tiny gold nanoparticles that originates from the anisotropic response of composing polycrystalline constituents [39].

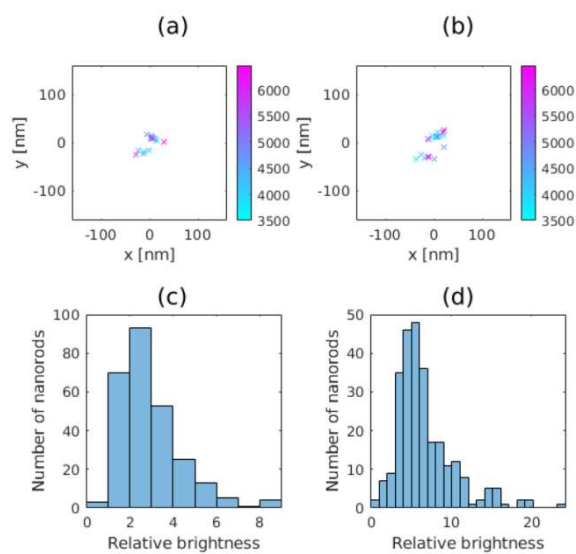

**Figure 7.** (a and b) dSTORM images of two Au nanorods labeled with Cy5 dye molecules, the color code shows the peak counts of localizations with enhanced brightness. (c) Mean and (d) maximum fluorescence enhancement values of ~260 nanorod clusters.

## 4. Conclusions

In our present study the synergy of the design and optimization of coupled fluorescent molecule - plasmonic nanoparticle systems, the controlled preparation and functionalization of plasmonic nanoparticles and the state-of-the-art dSTORM monitoring made it possible to prove that the level of fluorescence enhancement predicted theoretically is achievable experimentally.

Namely, (i) coupled Cy5 dye molecule and gold nanorod configurations optimal to enhance excitation and emission were designed moreover it was shown that the highest simultaneous excitation and emission enhancement can be ensured by tuning the plasmonic resonance peak in between the Cy5 dye molecule excitation and emission wavelengths, (ii) Au NRs with a desired geometry were synthesized, (iii) Cy5 dye molecules were aligned at a desired distance from the Au NRs' surface via hybridization of the Cy5-labelled oligonucleotide, (iv) dSTORM was capable of detecting individual molecules that are stochastically on-state around individual Au NRs, thus single on-state-molecule detection proved that an emission enhancement is reached on the level predicted theoretically for individual dipolar emitters at analogue distance from individual Au NRs. The explanation of the good agreement between theory and experiment is that only those Cy5 dye molecules were observable by dSTORM, which approximated the optimal configurations.

**Author Contributions:** Data curation, E.T. and T.N.; investigation, M.C.; methodology, D.U. and Gy.F.; software, B.B.; supervision, M.C.; validation, E.C. and M.E.; visualization, E.T., D.U. and T.N.; writing – original draft, D.U., T.N., Gy.F., E.C. and M.E.; writing – review and editing, M.C.

**Funding:** Coupled systems' numerical optimization and analyses has been supported by the European Union, co-financed by the European Social Fund EFOP-3.6.2-16-2017-00005 "Ultrafast physical processes in atoms, molecules, nanostructures and biological systems" and by the National Research, Development and Innovation Office (NKFIH) through the project "Optimized nanoplasmonics" (K116362). Super-resolution dSTORM imaging and data analysis were supported by the Hungarian Brain Research Program (2017-1.2.1-NKP-2017-00002) and the Multimodal Optical Nanoscopy (GINOP-2.3.2-15-2016-00036) project. The preparation and functionalization of plasmonic nanoparticles was supported by the National Research, Development and Innovation Office (NKFIH) through the project GINOP-2.3.2-15-2016-00038, FK 131446 and PD134282 as well as by the János Bolyai Research Scholarship of the Hungarian Academy of Sciences and the ÚNKP-19-4-SZTE-57 New National Excellence Program of the Ministry for Innovation and Technology (E. C.). The Ministry of Human Capacities, Hungary grant TUDFO/47138-1/2019-ITM is acknowledged.

**Conflicts of Interest:** The authors declare no conflict of interest.